\documentclass{article}

\usepackage[final, nonatbib]{nips_2017}

\usepackage[utf8]{inputenc} 
\usepackage[T1]{fontenc}    
\usepackage{hyperref}       
\usepackage{url}            
\usepackage{booktabs}       
\usepackage{amsfonts}       
\usepackage{nicefrac}       
\usepackage{microtype}      
\usepackage{graphicx}		
\usepackage{enumitem}
\usepackage{authblk}
\usepackage{caption}

\title{Automated flow for compressing convolution neural networks for efficient edge-computation with FPGA}

\author{
  \textbf{Farhan Shafiq}}
  \author{
  \textbf{Takato Yamada}}
  \author
  {\textbf{Antonio T. Vilchez}}
  \author{
  \textbf{Sakyasingha Dasgupta}}
\affil{LeapMind, Inc. \\ Tokyo, Japan \\ \{farhan, yamada, antonio, sakya\}@leapmind.io }

\begin{document}
\maketitle
\begin{abstract}
Deep convolutional neural networks (CNN) based solutions are the current state-of-the-art for computer vision tasks. Due to the large size of these models, they are typically run on clusters of CPUs or GPUs. However, power requirements and cost budgets can be a major hindrance in adoption of CNN for IoT applications. Recent research highlights that CNN contain significant redundancy in their structure and can be quantized to lower bit-width parameters and activations, while maintaining acceptable accuracy. Low bit-width and especially single bit-width (binary) CNN are particularly suitable for mobile applications based on FPGA implementation, due to the bitwise logic operations involved in binarized CNN. Moreover, the transition to lower bit-widths opens new avenues for performance optimizations and model improvement. In this paper, we present an automatic flow from trained TensorFlow models to FPGA system on chip implementation of binarized CNN. This flow involves quantization of model parameters and activations, generation of network and model in embedded-C, followed by automatic generation of the FPGA accelerator for binary convolutions. The automated flow is demonstrated through implementation of binarized "YOLOV2" on the low cost, low power Cyclone-V FPGA device. Experiments on object detection using binarized YOLOV2 demonstrate significant performance benefit in terms of model size and inference speed on FPGA as compared to CPU and mobile CPU platforms. Furthermore, the entire automated flow from trained models to FPGA synthesis can be completed within one hour.
%
%
\end{abstract}

\section{Introduction}
\begin{figure}
\begin{center}
\noindent\makebox[\textwidth]{\includegraphics[scale=0.55]{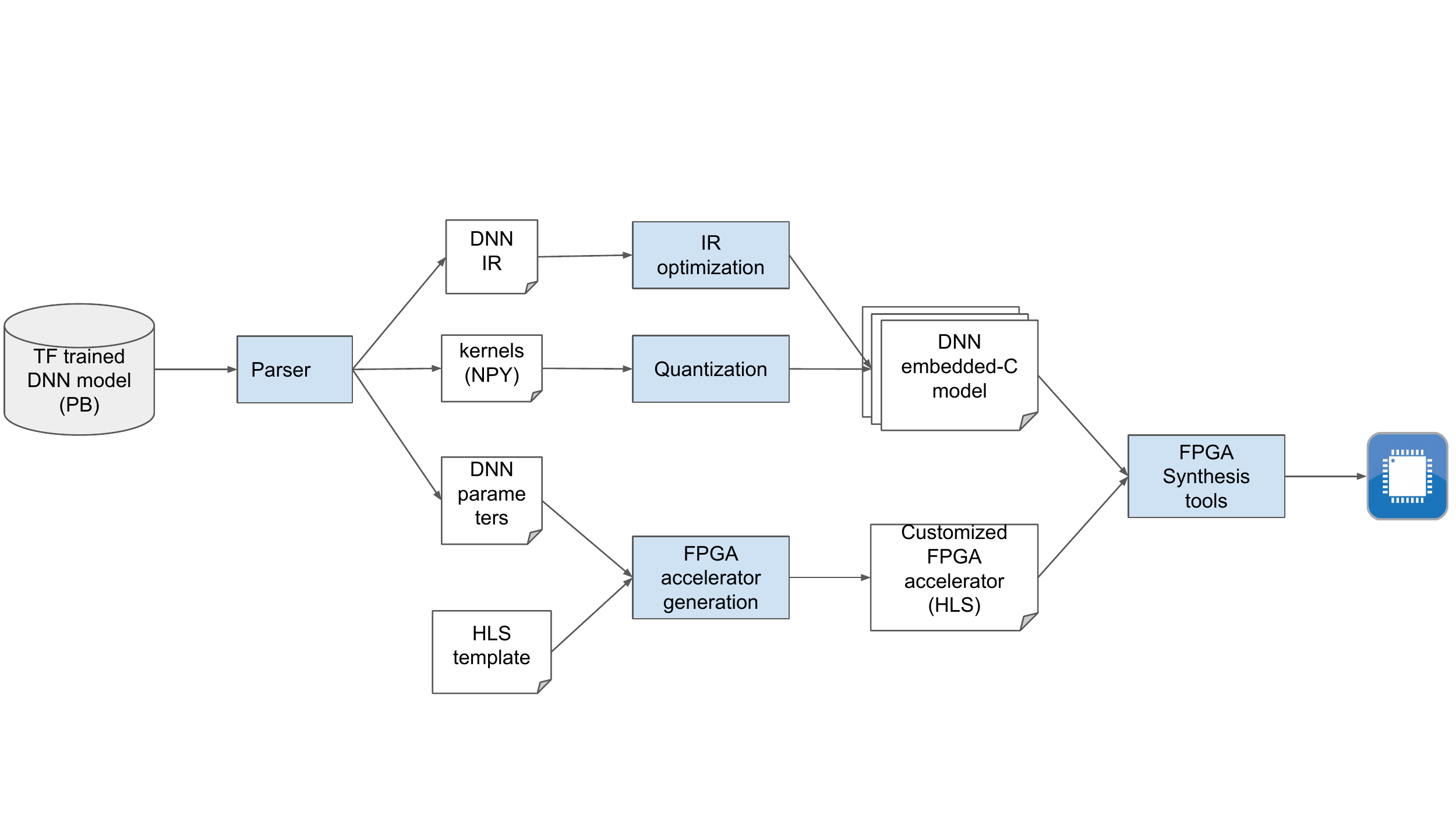}}
\caption{TensorFlow trained CNN model to FPGA implementation flow}
\label{fig1}
\end{center}
\end{figure}

%
%
Deep Convolutional Neural Networks (CNN) have achieved significant results in computer vision, speech recognition and language translation. However the computation and memory demands of recent CNN architectures require powerful GPUs, distributed CPU servers, Specialized ASIC or DSP processors. The size and power requirements of such platforms restrict the wide-spread adoption of CNN models for efficient edge computing, mobile devices and the Internet of Things in general.
Interestingly, recent results on compression of these models using a mix of techniques like pruning\cite{Deepcompression}, slimmed down architectures\cite{Mobilenets}, \cite{Squeezenet} and quantization to low bit-width, especially ternarized \cite{TWN} and binarized neural networks \cite{BNN} has shown that model size can be reduced dramatically while maintaining reasonable levels of accuracy. Furthermore, low bit-width CNN are particularly suitable for FPGA based acceleration due to bitwise operations involved in such models. As such quantized and compressed models can be effectively deployed on mobile devices based on such FPGA or SoC acceleration.
In this paper, we present an automatic flow from TensorFlow \cite{TF} trained CNN models to quantized CNN implementation on FPGA SoC, enabling efficient inference. Initially a full precision model is quantized to 1 bit weights and 2 bit activations, followed by retraining on the original dataset. Following this, the flow starts with the fully trained quantized model exported as Tensorflow protocol buffer format. This is followed by, the model being parsed and relevant graph transformations being applied. Embedded-C code for the quantized network is then generated and high level synthesis (HLS) implementation of the FPGA accelerator is customized for the quantized network. This is done using automated scripts that consider the model's memory requirements and computation complexity in order to choose the suitable level of parallelization and local memory usage. Figure ~\ref{fig1} depicts the block diagram of this flow. Currently the FPGA accelerator design is done with some input from humans rather than a full design space exploration. There have been a few other works in this area complementary to our approach, namely, Yaman et al. \cite{FINN}  presented automatic generation of FPGA accelerators from customizable Xilinx HLS templates. 
%
%

\section{Model parsing}
\begin{figure}
\begin{center}
\noindent\makebox[\textwidth]{\includegraphics[scale=0.4]{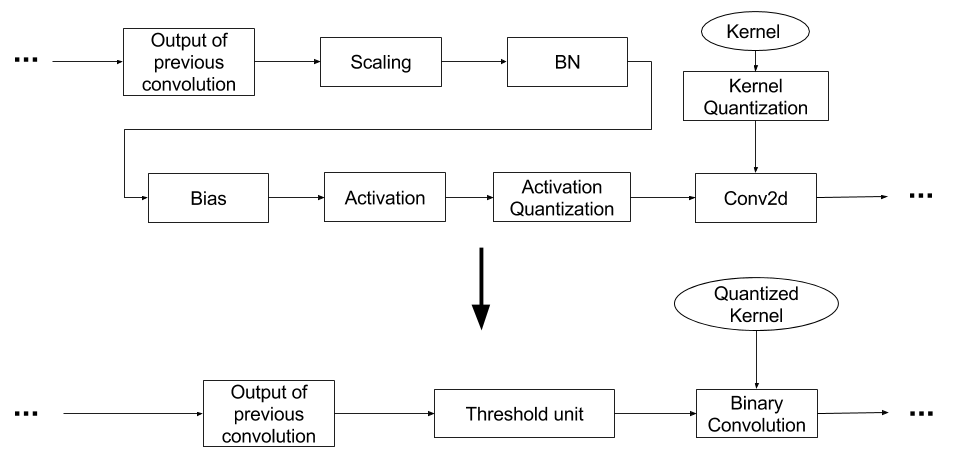}}
\caption{Subgraph between consecutive convolutions, replaced by a threshold unit.}
\label{fig2}
\end{center}
\end{figure}

%
%
The starting point of our framework is a fully trained model obtained from a specific deep-learning framework (e.g. TensorFlow, PyTorch, Caffe). In the case of TensorFlow, the model is first serialized into a binary protocol-buffer file. This contains both the computational graph and the model parameters. The key aspect of our framework is the automatic and transparent management of quantized activations and weights. If quantization is used during training the corresponding subgraphs will be pruned and replaced by their bit-wise counterparts, which are much more efficient for inference. Figure ~\ref{fig2}. shows an example of two consecutive convolutions with a subgraph that describes a linear operation between them. In this case the kernel quantization subgraph can be deleted. These subgraphs contain constant tensors of real numbers in 32-bit floating-point precision. However, as the training can be performed on the binary quantized CNN model, the weights in each layer can be packed efficiently. This allows a single 4 byte word to contain up to 32 weight values. This drastically improves the model size, enabling the weights to be represented as standard arrays in C-code. Furthermore, this effectively eliminates the additional time previously required for loading weights and enables the generation of compact, self-contained inference units.

\section{FPGA based acceleration}
%
%
\label{fpga}
Post embedded C code generation, potential parts for FPGA optimization are mapped to an FPGA accelerator. In general, CNN convolutional layers, especially binary convolutional layers are inherently suitable for parallelization and FPGA acceleration. The accelerator design space is limited, by the amount of computation resource on as well as the maximum data bandwidth available. Recent FPGA implementations of CNN accelerators address the memory bandwidth problem by allocating on-chip memory space for all kernels, inputs and outputs \cite{FINN}, thus minimizing off-chip communications. Such an implementation, although effective, requires large on-FPGA RAM blocks and consequently, expensive FPGAs. Hence accelerator parallelization approaches are heavily dependent on the target FPGA device’s computation resource, size of FPGA RAM blocks and the memory bandwidth.
\subsection{Strategy for parallelization}
%
%
We employ a scalable parallelization technique where the accelerator consists of the following building blocks.
\begin{itemize}[noitemsep, nolistsep, leftmargin=*]
\item \textbf{Processing Element (PE)} Multiple kernel elements are packed into a single word (32bit). In case of 1-bit kernel elements a single processing element (PE) can process 32 kernel elements in parallel. Each PE is followed by a 32-bit accumulator and it exploits intra-kernel parallelization.
\item \textbf{Processing Engine (PEN)} Multiple kernels can be processed in parallel. This can be achieved by employing a matrix of PEs processing the same input element and an element of different kernels in parallel. This exploits inter-kernel parallelism and increases input reuse.\footnote{It is also possible to exploit intra-input parallelization, however we do not discuss it here.} 
\end{itemize}

\subsection{Design assumption}
The FPGA accelerator is customized for the given network architecture and target FPGA device, in order to exploit the right parallelism. We assume that (1) Number of output feature maps is a multiple of 8. (2) Number of input feature maps is a multiple of 16 and (3) On-chip memory is limited. Assumptions (1) and (2) are in accordance with most of the popular network architectures in use while (3) is in accordance with the cost sensitive edge computing application. No assumptions are imposed on input/activation size.
\begin{figure}
\begin{center}
\includegraphics[scale=0.5]{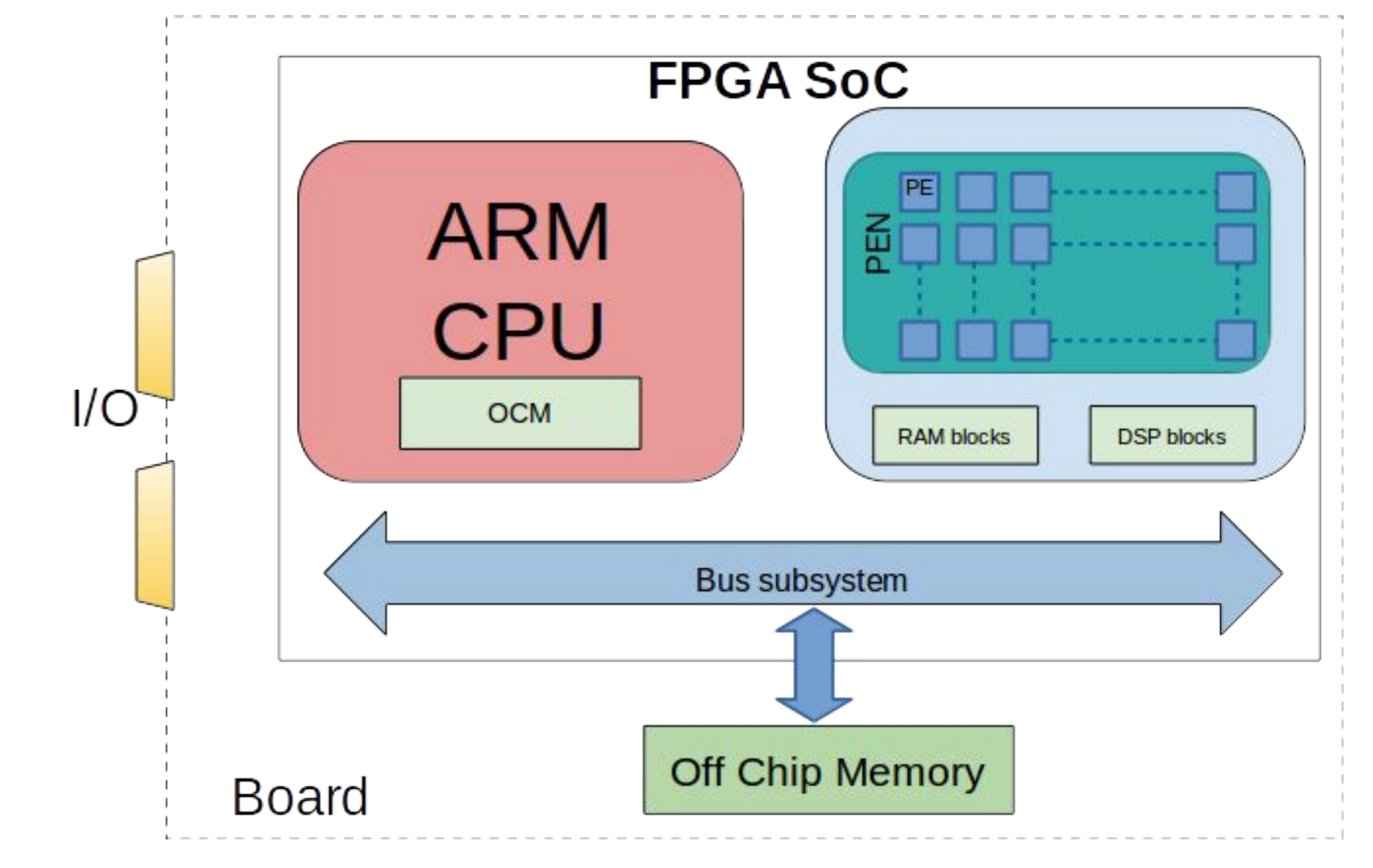}
\caption{FPGA SoC platform block diagram with parallel processing elements}
\label{fig3}
\end{center}
\end{figure}
\subsection{Accelerator generation}
The accelerator generation involves the following three steps:
(1) Customize the basic building block i.e. Processing Element ($PE$) depending on bits per element for kernel and input.
(2) Customize the Processing Engine ($PEN$) in the form of a matrix of ($PEs$). Number of ($PEs$) can be from 16 up to min($depth_i$) where $depth_i$ is the depth dimension of any layer’s input
(3) Automatically calculate other related parameters and control code. Figure~\ref{fig3} shows a system block diagram.
%
%
\begin{figure}
\centering
    \includegraphics[scale=0.5]{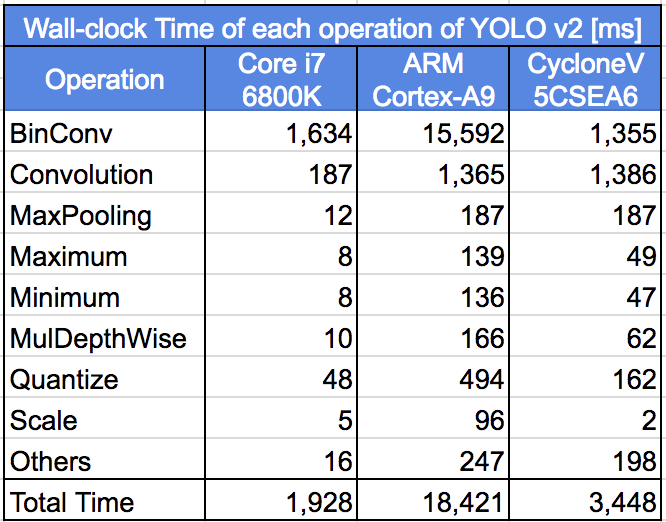}
    \caption{Wall-clock time (ms) for each operation of YOLO v2}
    \label{fig4}
\end{figure}
\subsection{Data order optimization}

Input, output and kernels for CNN can be visualized as 3-dimensional data arrays. In typical frameworks these arrays are stored in the order of $Depth \times Width \times Height$ (height dimension is updated first) or $Depth \times Height \times Width$. We propose to order the input, output and kernels in $Height \times Width \times Depth$ or $Width \times Height \times Depth$ (depth dimension is updated first) dimensions. This Depth-first ordering has a few merits as discussed below.

\subsection{Memory bandwidth optimization}
Memory bandwidth from FPGA device to off-chip DRAM is a usual performance bottleneck. This bottleneck can be relaxed by using burst transfers. However, the burst size is limited by the continuity of the memory addresses being accessed. The proposed ordering maximizes the burst size and improves effective memory bandwidth as a result. Figures below highlight the merit of depth first ordering.

\begin{figure}
\centering
\begin{minipage}{0.4\textwidth}
\centering
\includegraphics[width=\linewidth]{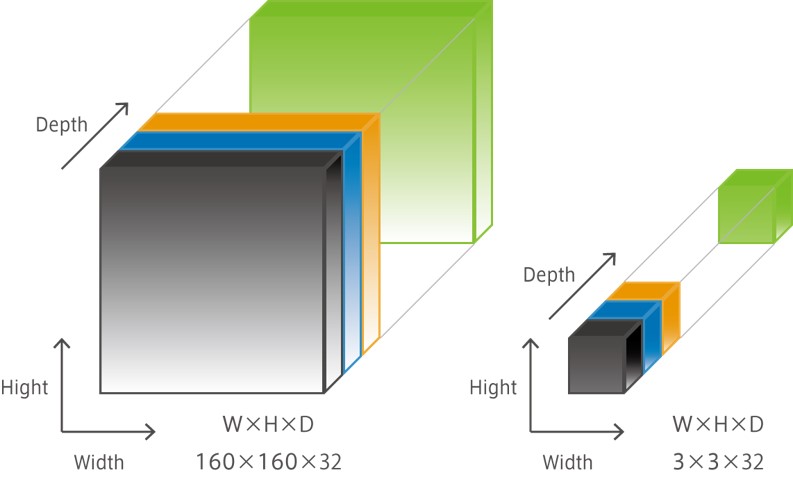}
\end{minipage}
\begin{minipage}{0.4\textwidth}
\centering
 \includegraphics[width=\linewidth]{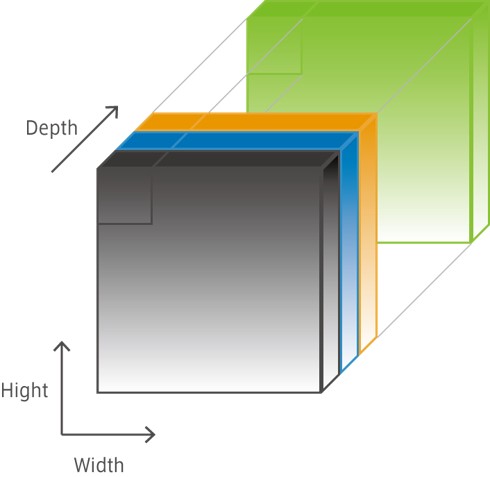}
\end{minipage}
\caption{Element-wise overlap of input and kernel }
\label{fig:sup1}
\end{figure}

\textbf{W-bar and D-bar: } \newline \\
Figure~\ref{fig:sup1} shows the element-wise overlap of input and kernel and respective sizes of W-bar and D-bar. Assuming that input and kernel consist of $Ih \times Id$ and $Kh \times Kd$ different width-wise arrays with $Iw$ and $Kw$ elements each respectively. These arrays are termed as “W-bar”. In contrast, "D-bar" signifies the $Ih \times Iw$ and $Kh \times Kw$ different depth-wise arrays with $Id$ and $Kd$ elements each respectively.

\begin{figure}
\centering
\begin{minipage}{0.9\textwidth}
\centering
\includegraphics[width=\linewidth]{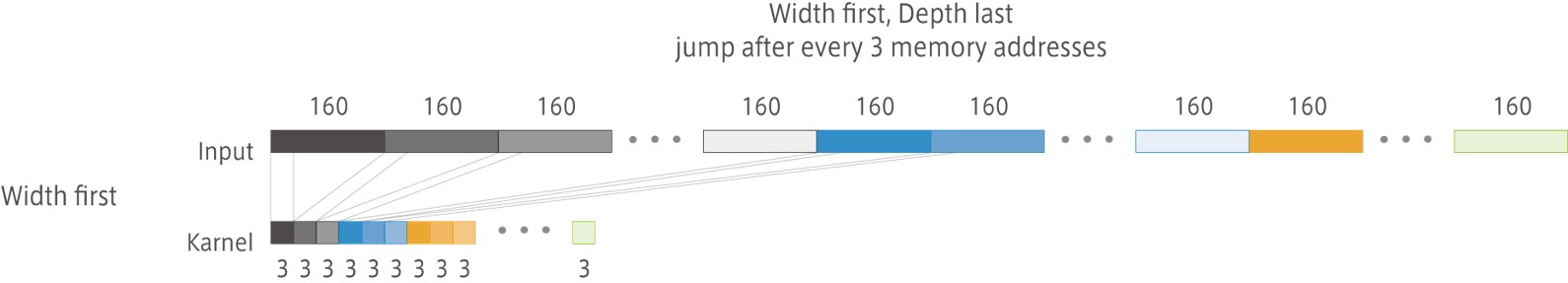}
\end{minipage}
\begin{minipage}{0.9\textwidth}
\centering
 \includegraphics[width=\linewidth]{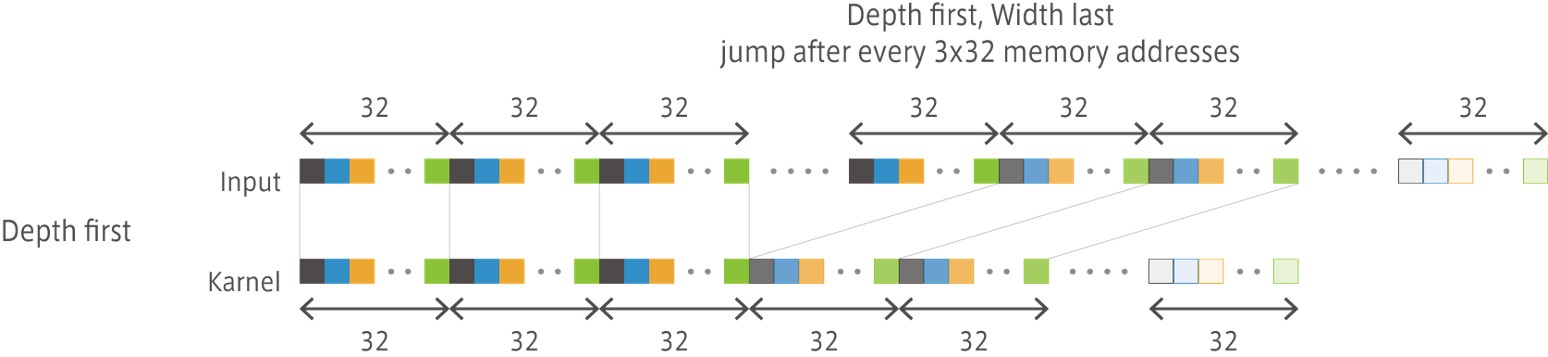}
\end{minipage}
\caption{Differences in memory access based on data-ordering }
\label{fig:sup2}
\end{figure}

\textbf{External and Local Memory access: } \newline \\
Figure~\ref{fig:sup2} depicts the difference of memory access continuity for width-first ordering as opposed to depth-first ordering. 

\begin{figure}[h]
\centering
\begin{minipage}{0.4\textwidth}
\centering
\includegraphics[width=\linewidth]{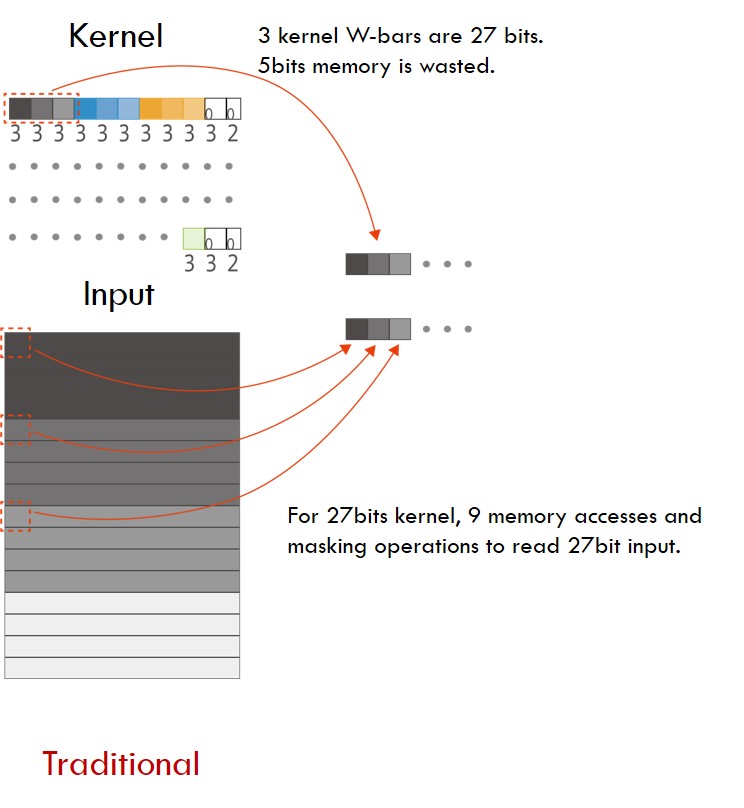}
\end{minipage}
\begin{minipage}{0.4\textwidth}
\centering
 \includegraphics[width=\linewidth]{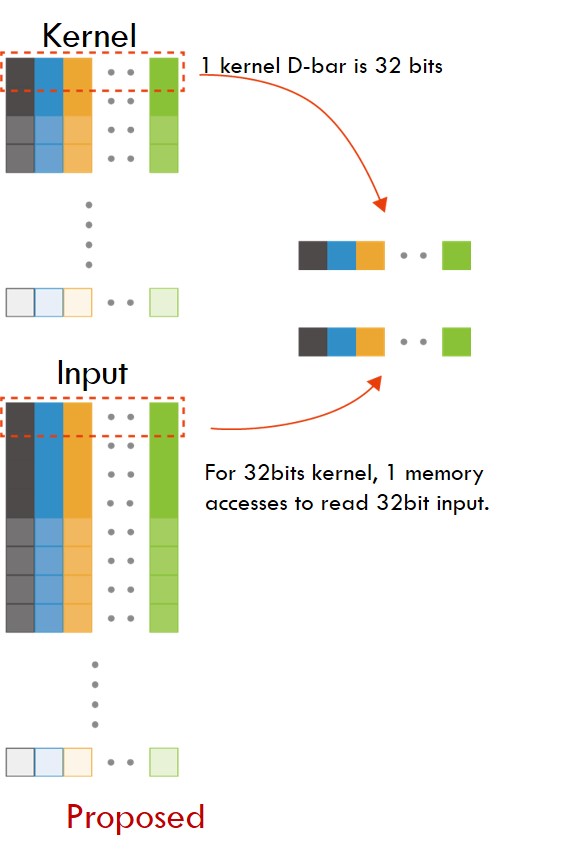}
\end{minipage}
\caption{Efficient implementation of bit-packing on local RAM blocks with proposed data-order optimization. }
\label{fig:sup3}
\end{figure}

\begin{itemize}
\item \textbf{Input from External memory:} The kernel W-bar overlaps with only Kw elements of the input W-bar at a time, which results in a jump in the memory addresses of the input when the next kernel W-bar is processed. In total over the course of a $Kw \times Kh \times Kd$ kernel, there are $Kh \times Kd$ jumps in the memory addresses of the input. On the other hand, the kernel D-bar overlaps with $Kd \times Kw$ elements of the input W-bar at a time, which results in only $Kh$ jumps over the course of a $Kh \times Kw \times Kd$ kernel resulting in a longer continuity in memory addresses and better burst performance.

\item \textbf{Local memory bit-packing and Processing Elements (PE):} As shown in Figure~\ref{fig:sup3}, The depth-wise ordering also results in a cleaner and efficient implementation of bit-packing for quantized values on the local RAM blocks, eliminating the need for multiple local memory accesses and masking operations. For example, assume that a single processing Element (PE) can process 32 binary inputs at a time. Proposed ordering helps avoid multiple memory access and bit-masking operations that would be necessary with traditional ordering. Moreover, the proposed ordering enables coarse-grain access of the local RAM (1 D-bar at a time as opposed to the fine-grain 1 W-bar or 1 byte at a time). This results in optimization of the local RAM access circuitry. 

\item \textbf{Inter-kernel parallelism: } Processing multiple kernels in parallel (inter-kernel parallelism) is an effective way to increase processing parallelism as well as increase input reuse. The same input is convolved with multiple kernels, getting output elements (at the same index) for multiple feature maps (output channels). With inter-kernel parallel processing, the order output elements are calculated, is naturally in the proposed depth first order. If the output is being saved on the local memory, proposed order enables a cleaner memory write circuitry. If these outputs are written directly to off-chip RAM, the proposed ordering enables better burst performance.
\end{itemize}

\begin{figure}[h]
\centering

\begin{minipage}{0.9\textwidth}
\centering
\includegraphics[width=0.5\linewidth]{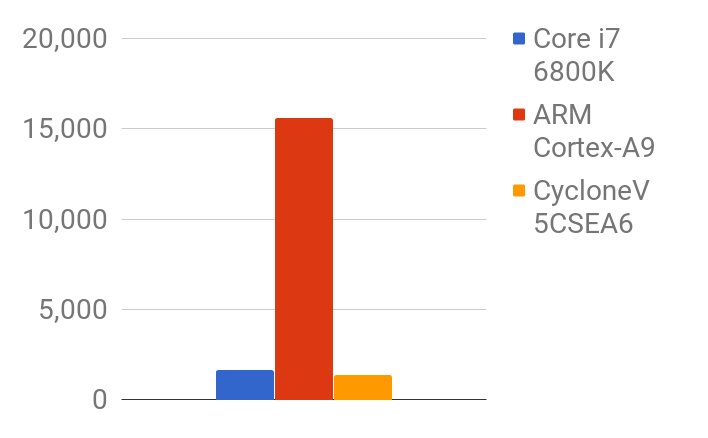}
\caption{Wall-clock time (ms) for binary convolution}
\label{fig5}
\end{minipage}
\begin{minipage}{0.9\textwidth}
\centering
 \includegraphics[width=0.5\linewidth]{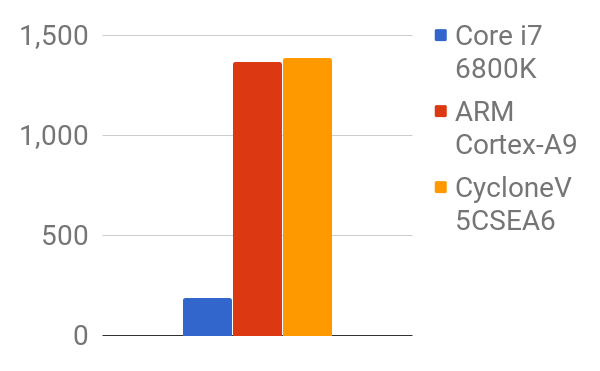}
 \caption{Wall-clock time (ms) for float convolution}
 \label{fig6}
\end{minipage}
\end{figure}
\section{Evaluation}
This section describes experimental environment setup and performance results from a benchmark CNN architecture.

We perform evaluation of our framework, by implementing YOLO v2 \cite{Yolo}. The network model mostly follows original Darknet-19 design but the input size is 320x320. The network was trained on PASCAL VOC 2012 using Tensorflow and converted into C program using our generation flow. "Weights" and "Activations" are quantized to 1bit and 2bit respectively(first and last layer are not quantized) resulting in a 32x smaller model (original Yolo v2 at 255.82 MB, compressed Yolo v2 at 8.26 MB). The generated C program is built for three separate cases: CPU(Core i7-6800K), Mobile CPU (ARM Cortex-A9), Mobile CPU with FPGA (Cyclone-V 5CSEA6 SoC containing ARM Contex-A9). Every runtime was compiled by g++ with –O3 option. 

In the FPGA-SoC case, binary convolution accelerator was generated using our customization and generation flow using Altera HLS compiler and synthesized as a hardware accelerator on the Programmable Logic. A comparison of performance on the three devices are reported. Figures ~\ref{fig4}, ~\ref{fig5} and ~\ref{fig6} show the wall-clock time of each operation running on each device.  In terms of binary convolution operation (BinConv-Core) which is accelerated by FPGA, Mobile CPU with FPGA case achieves up to a x11.50 and x1.21 speed up over only Mobile CPU and normal CPU case respectively. In total, the FPGA results 5.34 times faster than the case using only Mobile CPU but it’s 1.78x slower than corei7 CPU. The flow from a trained TF model to FPGA synthesis takes roughly around an hour.

\section{Conclusion}
This paper presented an automated flow from trained Tensorflow DNN models to
Binarized (quantized) FPGA SoC implementation targeting size, cost and power constrained
edge computing applications. An FPGA accelerator for quantized convolution is generated
with accordance to specific constraints. A data-ordering optimization is proposed for
improved memory performance. The proposed automated flow was evaluated based on
implementation of the state of the art YOLO-V2 object detection framework. The resulting
implementation achieves comparable performance to corei7 CPU. However, further
improvements are needed for real time applications. In future works, further improvements of
FPGA acceleration and SoC platforms will be explored.

\bibliographystyle{acm}
\bibliography{bibliography}


\end{document}